\documentstyle[12pt]{article}
\def \a{\alpha}
\def \b{\beta}

\def \ep{\epsilon}
\def \s{\sigma^1}
\def \e{\eta}
\def \o{\omega}
\def \p{\partial}
\def \d{\delta}
\def \su{\sum_{k}}
\def \dg{\dagger}
\begin{document}

\centerline{\bf Mode Analysis and  Duality Symmetry in Different Dimensions}

\vskip 0.5in

\centerline{R. Banerjee \footnote{rabin@boson.bose.res.in} and B. Chakraborty 
\footnote{biswajit@boson.bose.res.in}}
\centerline{S.N.Bose National Centre for Basic Sciences}
\centerline{Block-JD, Sector-III, Salt Lake}
\centerline{Calcutta-700091, India}

\vskip 1.5in

{\bf Abstract}

The problem of duality symmetry in free field models is examined in
details by performing a mode expansion of these fields which provides
a mapping with the purely quantum mechanical example of a harmonic oscillator.
By analysing the duality symmetry in the harmonic oscillator, we show that
the massless scalar theory in two dimensions display, along with the expected
discrete $Z_2$ symmetry, the continuous $SO(2)$ symmetry as well. The same
holds for the free Maxwell theory in four dimensions, which is usually
regarded to manifest only the $SO(2)$ symmetry. This leads to the new result
that, following a proper interpretation, 
the duality groups in two and four dimensions become identical.
Incidentally, duality in quantum mechanics is generally not covered
in the literature that considers only $D=4k$ or $D=4k+2$
spacetime dimensions, for integral $k$.

\pagebreak

\noindent{\bf I.Introduction}

The crucial role played by duality symmetry either in field or string
theories is becoming increasingly evident \cite{R}. As briefly
reviewed below, the conventional interpretation
of this symmetry leads to distinct groups in $4k$ or $4k+2$ dimensions
\cite{SS,DGHT}. This shows that there is a basic difference in the treatment of
duality symmetry in these dimensions. The motivation of
the present work is to show that it is possible to obliterate
this difference and reveal duality symmetry in a general
framework that encompasses both these dimensions. As a consequence, the
duality symmetry groups also become identical, irrespective of the
dimensionality. The explicit
calculations will be presented in the context of the Maxwell
theory in four dimensions and the free massless scalar theory in
two dimensions.

Historically, the source free Maxwell's equations were the first
to display the property of duality symmetry which involves a formal $SO(2)$
rotation, apart from a trivial scale factor, in the space of
electric and magnetic fields,
$$\pmatrix{{\bf E} \cr {\bf B}}\rightarrow \pmatrix{{\bf E}' \cr {\bf B}'}
= \pmatrix{\cos \theta & \sin \theta \cr -\sin \theta & \cos \theta}
\pmatrix{{\bf E} \cr {\bf B}} \eqno(1.1a)$$
or, equivalently a $U(1)$ transformation for the combination
$({\bf E}+i{\bf B})$
$$({\bf E}+i{\bf B})\rightarrow ({\bf E}'+i{\bf B}')=e^{-i\theta}
({\bf E}+i{\bf B}) \eqno(1.1b)$$
Using the language of differential two-forms $F$ and its dual
$\tilde F$, defined as,
$$F=E_i dx^0\wedge dx^i+ {1\over 2}F_{ij}dx^i \wedge dx^j$$
$${\tilde F}=-B_i dx^0\wedge dx^i 
+{1\over 2}{\tilde F}_{ij}dx^i\wedge dx^j\eqno(1.2)$$
with $B_i={1\over 2}\ep^{ijk}F_{jk}$ and 
$E_i=F_{0i}={1\over 2}\ep^{ijk}{\tilde F}_{jk} $ being the components
of the magnetic and electric fields respectively, (1.1a) is recast as,
$$\pmatrix{F \cr {\tilde F}}\rightarrow \pmatrix{F' \cr {\tilde F}'}
= \pmatrix{\cos \theta & -\sin \theta \cr \sin \theta & \cos \theta}
\pmatrix{F \cr {\tilde F}} \eqno(1.3)$$
As is well known this is a symmetry of the equations of motion
$dF=0$ and $d\tilde F=0$ only, but not of the action $S=\int
d^4x tr(FF-\tilde F\tilde F)$. Incidentally, this analysis is
generic for any abelian $N=2k$-form 
fields in $D=4k$ dimensions, for integral $k$.

The corresponding situation in two dimensions (which is generic
for $D=4k+2$ dimensions, for integral $k$) has also been
studied. In the case of the free massless scalar field (which
can be regarded as a zero form potential) in two dimensions,
the equations of motion are invariant
under $Z_2 \times SO(1, 1)$ transformations, although the action
is not. This difference from the four dimensional example is
attributed to the basic identities governing the dual operations,
$${\tilde {\tilde F}}=-F ;D=4k$$
$${\tilde {\tilde F}}=F ;D=4k+2\eqno(1.4)$$
To elevate the duality at the level of the action, it was
naturally imperative to define the relevant transformations in
terms of the basic variables which are the associated
potentials rather than the field tensors. This is possible by
rewriting the action in terms of two potentials. Incidentally,
the introduction of a second potential $A'$ is essentially tied to
the fact that the dual field $\tilde F$ is closed by the
equation of motion, so that one can write ${\tilde F}=dA'$ as an on-shell
relation. It was also shown that the duality groups $G$
preserving the invariance of the action were the subgroups of
those found earlier that preserve the invariance of the
equations of motion. In fact the former was obtained by taking an
intersection with $O(2)$, the group of invariance
of the energy-momentum tensor $(T_{\mu \nu}\sim (F_{\mu}F_{\nu}+
F_{\nu}F_{\mu}))$(here the unwritten indices have been summed over).
Specifically, these were \cite{SS,DGHT}
$$G=SO(2);D=4k$$
$$G=Z_2; D=(4k+2) \eqno(1.5)$$
It is clear, therefore, that  a fundamental difference is
observed in the study of
duality symmetry in $4k$ and $4k+2$ dimensions.

To put the  above discussion in a proper perspective, it might
be useful to mention that the original study of duality symmetry
in the context of the equations of motion can be understood in
an alternative way that does not involve these equationa at all.
Indeed, it is simple to check 
that by only demanding 
the invariance of the dual operation $F\rightarrow\tilde F$
under some 
transformation (like (1. 3)) yields the $SO(2)$ group for four dimensions. 
Consider, for instance, the following transformation,
$$
\pmatrix {F \cr {\tilde F}}\rightarrow \pmatrix {F' \cr {\tilde F}'}
= \pmatrix {p & q \cr r & s}\pmatrix {F \cr {\tilde F}}\eqno(1.6)
$$ 
If we demand that ${\tilde F}'$ is indeed the dual of $F'$, then it follows
that $p=s$ and $q=-r$. Hence, upto a trivial scale factor, the transformation
matrix in (1.6) can be identified with the standard $SO(2)$ matrix (1.3).
The same logic holds for two
dimensions also where the relavant group is found to be
$Z_2\times SO(1, 1)$ instead of $SO(2)$. Hence the study of
duality symmetry truly becomes meaningful only with
regard to the respective actions.

A natural question that may arise here concerns the fate of
quantum mechanics, which can be regarded as a field theory in $0+1$
dimension ($D=1$), as this is not covered by the general equations (1.4,1.5).
Likewise, because of dimensional reasons, the form based 
analysis done earlier is also inapplicable.
The fact that 
both the free Maxwell theory and the $2$-dimensional
massless scalar theory reduce to an assembly of infinite number
of harmonic oscillators (HO), as can be seen through a mode expansion,
lends further credence to the study of duality symmetry
in the quantum mechanical context. 
One can thus assign a more fundamental status to the corresponding
Fourier transformed amplitudes, as these amplitudes, albeit complex,
undergo harmonic oscillations. After all, the particle content of 
such free field theories are identified with the corresponding excitations
in various modes.
Interpreted in this fashion,
there seems to be a mapping of the results for the Maxwell and
scalar field theories and the noted difference in the duality
groups is clearly not the complete story.

In this paper we explicitly show how the HO manifests a duality symmetry.
The corresponding duality group is $SO(2)$. The method of deriving a
duality symmetric lagrangian for the HO is easily generalisable to higher
dimensions; in particular this derivation is given for the scalar and
Maxwell theories in two and  four dimensions respectively.

The results of the HO analysis are next
put in a proper perspective with regard to the field theoretical models.
As already stated, the free fields can be thought of as an assembly
of an infinite number of HOs. Hence the feature of duality symmetry in
free field theories should be understandable from the analogous phenomenon
for the HO. At this point one is led to an impasse since the duality group
for a free scalar theory in two dimensions is known to be $Z_2$, contrary to
the $SO(2)$ group for the
HO. We resolve this apparent paradox by performing a detailed mode analysis
in various theories. In this context it becomes desirable to
consider the `complex' HO instead of the real one. The former,
in contrast to the latter, manifests either the $Z_2$ or $SO(2)$
symmetry by a suitable redefinitions of variables. A one to one
correspondence between the modes in the free fields and the
complex HO is then established. From this correspondence
it is shown that the scalar
theory manifests both the expected $Z_2$ as well as the $SO(2)$
symmetry, depending on
the interpretation of the results. Similar conclusions also hold
for the Maxwell theory. It may be recalled, however, that the conventional
interpretation of duality in the Maxwell theory admits only the $SO(2)$
symmetry.

The paper is split into five sections. In section II, duality symmetry in the
HO both for real and complex variables, is analysed. Sections III and IV
describe the corresponding analysis
for the scalar and Maxwell theories, including the comparison with the HO
formulation. Section V contains the concluding remarks. An appendix is
included to derive the explicit form of the duality generator
corresponding to the $k-th$ mode and also to
illuminate how the HO itself can be cast in the
electromagnetic form.

\vskip 0.5in

\noindent{\bf II.The  one-dimensional Harmonic Oscillator}

\vskip 0.3in

Consider a one dimensional harmonic oscillator given by the lagrangian
$$L={1\over 2}({\dot q}^2-\o^2q^2)\eqno(2.1)$$
It is possible to introduce an electromagnetic notation and rewrite (2.1) as
$$L={1\over 2}(E^2-B^2)\eqno(2.2)$$
with $E={\dot q}$ and $B=\o q$. With this, the equation of motion
$${\ddot q}+\o^2q=0\eqno(2.3)$$
along with the `Bianchi identity' $({\dot B}=\o E)$ can be expressed
compactly as,
$${\p \over {\p t}}(E+iB)=i\o (E+iB)\eqno(2.4)$$
which has a manifest $U(1)$ dual symmetry $(E+iB)\rightarrow e^{i\phi}(E+iB)$,
with $E$ and $B$ regarded as independent variables. However this is not a
symmetry of the Lagrangian, as can be easily seen from (2.2). But our objective
is to construct a Lagrangian which is equivalent to the former, 
and enjoys this duality
symmetry. The basic idea \cite{BW} is to linearise (2.1) by invoking an
additional variable `$p$' in an enlarged configuration space as,
$$L={1\over 2}[\o (p{\dot q}-q{\dot p})-\o^2(q^2+p^2)]\eqno(2.5)$$
Here a symmetrisation of the kinetic term has been performed.
As can be easily seen, the equation of motion for $p$ ($\o p={\dot q}$),
when substituted in (2.5), yields (2.1). It just happens in this case
that $p$ is the momentum conjugate to $q$. But this need not be true
always. In fact we shall regard $p$ just as an additional variable in an
enlarged two dimensional configuration space, as we have mentioned earlier.

By labelling $q=x_2$ and $p=x_1$,  (2.5) takes the form,
$$L_{+}={1\over 2}\o \ep_{\a \b}x_{\a}{\dot {x_\b}}
-{1\over 2}\o^2x_{\a}x_{\a}\eqno(2.6a)$$
On the other hand, labelling in the reverse order,i.e. $q=x_1$ and $p=x_2$,one
gets,
$$L_{-}=-{1\over 2}\o \ep_{\a \b}x_{\a}{\dot {x_\b}}
-{1\over 2}\o^2x_{\a}x_{\a}\eqno(2.6b)$$
Compactly, one can write,
$$L_{\pm}(X)=\pm {1\over 2}\o X^T\ep {\dot X}-{1\over 2}\o^2X^TX .\eqno(2.7)$$
Here we have used a  matrix notation. Thus $X$ stands for the doublet
$\pmatrix{x_1 \cr x_2}$ and $\ep$ for the $2\times 2$ matrix
$\pmatrix{0 & 1\cr {-1}& 0}$. Note that the cherished aim of obtaining a
duality symmetric lagrangian has been fulfilled. The above lagrangians
$L_{\pm}(X)$ are invariant under $SO(2)$ duality rotation
$X\rightarrow R^{+}X$ or equivalently, $E\rightarrow R^{+}E$,
$B\rightarrow R^{+}B$ in the electromagnetic symbols. Furthermore, the two
Lagrangians $L_{+}$ and $L_{-}$ are
swapped into each other $(L_{+}\leftrightarrow L_{-})$ by an improper $O(2)$
rotation ($X \rightarrow R^{-}X$). Proper (Improper) rotations
will be designated by $R^+(R^-)$ in the rest of the paper.
The generator of the $SO(2)$ duality rotation, in $L_{+}$ for example,
is given by,
$$G=-{1\over 2}\o X^TX$$
satisfying $\d X=\theta \{X, G\}$, for an infinitesimal $SO(2)$ rotation
$\theta $. This can be easily verified from the symplectic
bracket following from (2.7),
$$\{x_{\a },x_{\b }\}={1 \over \o }\ep_{\a \b}$$
The lagrangians $(L_{\pm})$ are refered to as chiral Lagrangians, as the
corresponding ``angular momentum''(the rotational generator in the internal
$x_1x_2$ space)
$$J_{\pm}=\pm {1\over \o}H= \pm {1\over 2}\o X^TX \eqno(2.8)$$
has positive or negative eigenvalues. Here $H$ is the Hamiltonian common
to both $L_{\pm}$ (2.7). In this sense (2.7) represents the dynamics of
chiral oscillators.

It is now an interesting and instructive exercise to show the combination
of the left and right chiral oscillators to reproduce the Lagrangian of
the usual HO. This
is done by following the soldering technique suggested in \cite{S} and
developed fully in \cite{ADW,ABW}.

Consider therefore $L_{+}(X)$ and $L_{-}(Y)$, given
as functions of two independent variables $X$ and $Y$, respectively.
Under a transformation  which preserves the difference $(X-Y)$,
i.e.
$$\d X= \d Y =\e \eqno(2.9)$$
the Lagrangians $L_{\pm}(Z)$ transform as,
$$\d L_{\pm}(Z)=\ep_{\a \b}\e_{\a}J_{\b}^{\pm}(Z)=\e^T \ep J^{\pm}(Z);Z=(X,Y)\eqno(2.10)$$
where ,
$$J^{\pm}(Z)=\pm {\dot Z}+ \o \ep Z\eqno(2.11)$$
Introduce a variable $B= \pmatrix{B_1 \cr B_2}$, that will effect the
soldering, transforming as,
$$\d B=-\ep \e \eqno(2.12)$$
Eventually, by a series of iterative steps, it can be shown that the
soldered Lagrangian,
$$L^s=L_{+}(X)+L_{-}(Y)-B^T(J^{+}(X)+J^{-}(Y))-B^TB \eqno(2.13)$$
is invariant under the complete set of transformations (2.9) and (2.12).

The equation of motion for the auxiliary field $B$,
$$B=-{1\over 2}(J^{+}(X)+J^{-}(Y))\eqno(2.14)$$
which follows from (2.13), when substituted back in the same equation, yields,
$$L^s={1\over 2}({\dot Z}^T{\dot Z}- {\o}^2Z^TZ) \eqno(2.15)$$
where $L^s$ is no longer a function of $X$ and $Y$ independently, but
only of their difference,
$$Z={1\over {\sqrt 2}}(X-Y) \eqno(2.16)$$
The above Lagrangian characterises a bi-dimensional oscillator. It is
duality symmetric under the complete $O(2)$ transformation
$Z \rightarrow R^{\pm}Z$. It is straightforward to show from (2.14) that
the transformation law (2.12) is properly reproduced, thereby serving as a
consistency check on the soldering programme.

Let us next consider the example of the `complex' HO. They occur naturally
as the Fourier modes of several free field theories and thus will be useful
for the subsequent analysis. Besides, this
is an instructive example where distinct variable redefinitions are possible
which show a reversal of roles of the duality transformations. Consider,
therefore, the following Lagrangian,
$$L={1\over 2}({\dot \phi}^*{\dot \phi} - \o^2 \phi^*\phi )\eqno(2.17)$$
Linearising the above Lagrangian, by introducing additional variables
$\pi$ and $\pi^*$ in an enlarged configuration space, one gets
$$L={1\over 2}\o(\pi^*{\dot \phi}+ \pi {\dot \phi}^*)
-{1\over 2}\o^2 (\pi^* \pi +\phi^*\phi ) \eqno(2.18)$$
Labelling $\phi =q_{1}$ and $\pi =q_{2}$ and then again in the reverse
order i.e. $\phi =q_{2}$ and $\pi =q_{1}$, the following
``chiral'' forms of the Lagrangian are obtained,
$$L_{\pm}(Q)=\pm {1\over 2}\o Q^{\dg}\ep {\dot Q}
-{1\over 2}\o^2Q^{\dg}Q \eqno(2.19)$$
with $Q=\pmatrix{q_{1}\cr q_{2}}$.

The occurence of the $\ep $ matrix indicates that, as before, the Lagrangians
$L_{\pm}$ are invariant under the $SO(2)$ transformation
$Q\rightarrow R^{+}Q$. Similarly, the improper rotations $R^{-}$ 
induce a swapping $L_{+} \leftrightarrow L_{-}$.
Exactly in analogue with the real HO, the generator of duality rotation
is found to be,
$$G=-{1\over 2}\o Q^{\dg }Q$$

Now consider the following alternative way of relabelling the $(\phi ,\pi )$
variables in (2.18),
$$\phi =q_1 ; \pi =iq_2 \eqno(2.20a)$$
and then as
$$\phi =q_2 ; \pi =-iq_1 \eqno(2.20b)$$
which yields the following structures for the Lagrangians,
$$L_{\pm}(Q)={1\over 2}(\pm i\o {\dot Q}^{\dg}\s Q-\o^2 Q^{\dg}Q)\eqno(2.21)$$
where  $\s=\pmatrix{0 & 1\cr 1 & 0}$ is the first Pauli matrix.
These Lagrangians are invariant under a discrete $Z_2$ transformation
$Q\rightarrow \s Q$, while the swapping $L_{+}\leftrightarrow L_{-}$ is
effected by $Q\rightarrow \ep Q$. Clearly therefore, the roles of proper
and improper rotations are reversed from the previous case. Compared to
the real example, the complex HO has a richer symmetry structure that is
essentially tied to the complex nature of the variables, allowing for
alternative redefinitions.

To complete the analysis, the soldering of the complex `chiral' oscillators
(2.19) is done to reproduce the complex HO. Consider a transformation,
$$W \rightarrow W'=W +\e \eqno(2.22)$$
Under this, (2.19) transforms as,
$$\d L_{\pm}(W)=\e^{\dg}\ep J^{\pm}-{J^{\pm}}^{\dg}\ep \e ;W=(Q,R) \eqno(2.23)$$
where,
$$J^{\pm}(W)=\pm {1\over 2}\o {\dot W}+{1\over 2}\o^2\ep W \eqno(2.24)$$
is the counterpart of (2.11).

Introduce a (column) matrix-valued variable $B$ transforming as,
$$\d B =-\ep \e \eqno(2.25)$$
Then the first iterated Lagrangian $L_{\pm}^{(1)}$, defined as,
$$L_{\pm}^{(1)}=L_{\pm}-B^{\dg}J^{\pm}-J^{\pm\dg}B \eqno(2.26)$$
can be shown to transform as,
$$\d L_{\pm}^{(1)}=\mp {1\over 2}\o (B^{\dg}{\dot \e}+{\dot \e}^{\dg}B)
-{1\over 2}\o^2(B^{\dg}\ep \e-\e^{\dg}\ep B) \eqno(2.27)$$
Assuming that $B$ does not depend on $W$, one finds,
$$\d L^{(1)}_{+}(Q)+\d L^{(1)}_{-}(R)=-\o^2(B^{\dg}\ep \e_
-\e^{\dg}\ep B) \eqno(2.28)$$

Just as in the previous section, the soldered Lagrangian $L^s$ is defined
as, 
$$L^s =L^{(1)}_{+}(Q)+L^{(1)}_{-}(R)-\o^2B^{\dg}B \eqno(2.29)$$
Using (2.25) and (2.28), one can easily show that $L^s$ is invariant 
$$\d L^s=0 \eqno(2.30)$$
under the above transformations (2.22) and (2.25). Eliminating the auxilliary
variables $B$ and $B^{\dg}$ from (2.29), by using the corresponding
equations of motion, one obtains
$$L^s={1\over 2}({\dot S}^{\dg}{\dot S}-\o^2S^{\dg}S)\eqno(2.31)$$
where,
$$S={1\over {\sqrt 2}}(Q-R)\eqno(2.32)$$
is a `gauge invariant' combination of variables.
Thus starting with chiral forms of Lagrangians $L_{+}(Q)$ and $L_{-}(R)$,
given as functions of $Q$ and $R$ respectively, we have constructed a
soldered Lagrangian $L^s$, which is a function of the difference 
$S$ (2.32) only. Thus the bi-dimensional complex HO is
manifestly invariant under the simultaneous
transformation, $\d Q=\d R=\e $. Incidentally, the same conclusions are
obtained if one starts from (2.21) instead of (2.19).

Using these concepts, the duality symmetry in the context of free field
theories is better understood, as evolved in the subsequent sections.

\vskip 0.5in

\noindent{\bf III.Massless scalar fields in $1+1$ dimensions}

\vskip 0.5cm
The HO is quite ubiquitious in field theoretical models.
This is because a large number of free field models
can be thought of as an assembly of infinite number of free HOs,
each designated by the mode vector $\bf k$. In this section, we
shall carry out the mode analysis of the massless scalar fields in $(1+1)$
dimension and study the duality symmetry through these modes, simultaneously
revealing the close connection with the HO analysis carried out in the
previous section.

The Lagrangian of the model is given by,
$$L={1\over 2}\int dx({\dot \phi}^2(x)- \phi'^2(x)) \eqno(3.1)$$
Putting the system in a box of length $L$, one can make the Fourier 
decomposition of the real scalar field $\phi (x)$ as,
$$\phi (x)={1\over {\sqrt L}}\su e^{ikx}\phi_k (t) \eqno(3.2)$$
Here $k$ represents the space component of a 2-vector $k^{\mu}$, satisfying 
$k^{\mu}k_{\mu}=\o_k^2-k^2=0$ and $\phi_k^*=\phi_{-k}$. Substituting (3.2)
in (3.1), one gets
$$L=\su L_k \eqno(3.3a)$$
with
$$L_k={1\over 2}({\dot \phi}^*_k {\dot \phi}_k -
\o_k^2 \phi^*_k \phi_k)\eqno(3.3b)$$
representing a "complex" HO for the $k$-th mode (see (2.17)), as
$\phi_k$ is a
complex number in general. Thus one can proceed just as in the preceeding
section to linearise the Lagrangian and then relabel the variables in the
appropriate manner to obtain the following duality invariant
forms of the Lagrangian,
$$L_{k\pm}(Q_k)=\pm {1\over 2}\o_kQ_k^{\dg}\ep {\dot Q}_k
-{1\over 2}\o_k^2Q^{\dg}_kQ_k \eqno(3.4a)$$
and
$$L_{k\pm}(Q_k)={1\over 2}(\pm i\o_k {\dot Q_k}^{\dg}\s Q_k
-\o_k^2 Q_k^{\dg}Q_k)\eqno(3.4b)$$
with $Q_k=\pmatrix{q_{1k}\cr q_{2k}}$. Note that these expressions are
just (2.19) and (2.21), but with
only an additional subscript $k$-the mode index. It is clear that while
the duality group is $SO(2)$ for (3.4a), it is $Z_2$ for (3.4b). Recall
that, expressed in terms of the original scalar fields, only the latter
is manifested \cite{SS,DGHT,BW}.

We can now proceed with the soldering of these two Lagrangians
$L_{k+}(Q)$ and $L_{k-}(R)$, for two independent variables $Q$ and $R$,
as we have done in the previous section to finally get
$$L_k^s={1\over 2}({\dot S}_k^{\dg}{\dot S}_k-\o_k^2S_k^{\dg}S_k)\eqno(3.5a)$$
where,
$$S_k={1\over {\sqrt 2}}(Q_k-R_k)\eqno(3.5b)$$
is the `gauge invariant' combination of variables $Q_k$ and $R_k$.
Note that the above result follows irrespective of whether one starts from
(3.4a) or (3.4b). The soldered Lagrangian, which is just the expression
for the $k$-th mode, is thus manifestly invariant under the simultaneous
transformation, $\d Q_k=\d R_k=\e_k$. At this stage we can sum over
all the modes to get the complete soldered Lagrangian $L^s$ as,
$$L^s=\su L_k^s={1\over 2}\su ({\dot S}_k^{\dg}{\dot S}_k
-\o_k^2S_k^{\dg}S_k)\eqno(3.6)$$
Using the inverse Fourier transform, this can be easily shown to yield
$$L^s={1\over 2}\int dx \p_{\mu}S^{\dg}(x)\p^{\mu}S(x) \eqno(3.7)$$
where
$$S(x)={1\over {\sqrt L}}\su e^{ikx}S_k(t)\eqno(3.8)$$
is a doublet of real scalar fields. This is again given in terms of the
difference,
$$S(x)={1\over {\sqrt 2}}(Q(x)-R(x))\eqno(3.9)$$
where $Q(x)$ and $R(x)$ are obtained from $Q_k$ and $R_k$ using expressions
similar to (3.8).

On the other hand, as shown in \cite{BW}, the original model (3.1) can be
reexpressed, after a suitable redefinition of variables, in a linearised form
as,
$${\cal L}_{\pm}(\Phi)={1\over 2}(\pm {\dot \Phi}^T\s {\Phi}'
-{\Phi}'^T{\Phi}')\eqno(3.10)$$
where $\Phi =\pmatrix{\phi_1 \cr \phi_2}$. The  matrix swapping
${\cal L}_{+}\leftrightarrow {\cal L}_{-}$ is $\ep$. Again as shown in 
\cite{BW},
the soldering of ${\cal L}_{+}(Q)$ and ${\cal L}_{-}(R)$, where $Q$ and $R$
denote the independent fields corresponding to the positive and negative
components of the Lagrangian given in (3.10), yields,
$${\cal L}^s={1\over 2}\p_{\mu}S^{\dg}\p^{\mu}S \eqno(3.11)$$
where $S$ is identical to (3.9). Note that this is precisely
the Lagrangian density appearing in (3.7). This shows that writing the 
original model in the chiral form and then soldering, as in \cite{BW}, yields
the same result as the one obtained by first making a Fourier decomposition
(3.2) and then expressing this Lagrangian (a ``complex HO'') 
in a linearised chiral oscillator form $L_{k\pm}$ (3.4),next soldering
to get $L_k^s$ (3.5), followed by a final summation over all the modes to
get (3.7).
%

It may be recalled that (3.10) is the conventional form of the duality
symmetric action in two dimensions \cite{SS,DGHT,BW}.
Nevertheless, expressed in terms of its
modes, the massless scalar theory (3.3b) gets mapped to the complex HO,
thereby  manifesting either the $Z_2$ or the $SO(2)$
symmetry depending on the variable redefinitions. To establish compatibility
with (3.10) where only the $Z_2$ symmetry is revealed, recall
that (3.10) was obtained \cite{BW} by rewriting (3.1) in its linearised version,
$${\cal L}={1\over 2}(P{\dot {\phi}}-{\dot P}\phi -P^2-{\phi'}^2)\eqno(3.12)$$
where $P$ is an additional variable in an extended configuration space.
In order to get the form
${\cal L}_{+}$(3.10) for example, one has to make the following relabelling
$$\phi=\phi_1 $$
$$ P=\phi'_2 \eqno(3.13a,b)$$
Incidentally, the existence of the second scalar field $\phi_2(x)$ is
understood in the following manner. Since $\phi (x)=\phi_1(x)$ can be
regarded as a zero-form potential, the field one-form
$$F=d \phi_1=({\dot \phi}_1dt+ \phi_1'dx)$$
has the dual
$${\tilde F}=-({\dot \phi}_1dx + \phi_1'dt)$$
which is closed on-shell, so that in the absence of any nontrivial topology
 it must be exact. In other words there exists another
function $\phi_2(x)$ satisfying ${\tilde F}=-d \phi_2$. As can be easily
seen, here $P={\dot \phi}_1={\phi }'_2$.

To get the form ${\cal L}_{-}$, the relabelling has to be done in the reverse
order i.e. $(\phi=\phi_2;P=\phi'_1)$. In the rest of this section, we shall
only consider ${\cal L}_{+}$ for convenience.

At this stage, we can Fourier analyse the field $\phi_{\a}(x)$ ($\a=1,2$)
and $P(x)$ as,
$$\phi_{\a}(x)={1\over {\sqrt L}}\su e^{ikx}\phi_{\a k}(t)\eqno(3.14)$$
$$P(x)={1\over {\sqrt L}}\su e^{ikx}k \pi_k(t) \eqno(3.15)$$
We can easily see that (3.13b) implies, in terms of momentum space variables,
$$\pi_k=i\phi_{2k}\eqno(3.16)$$
Again, reality of $\phi_{\a}(x)$ implies,
$$\phi^*_{\a k}= \phi_{\a (-k)}$$
$$\pi^*_k=-\pi_{-k}\eqno(3.17a,b)$$
One can then proceed, as for the model (3.1), to obtain two equivalent forms
for $L_k$, starting from  (3.12). Using the Fourier decomposition
of both  $\phi $ and $P$ fields, we get,
$$L_k={1\over 2}[k (\pi_k{\dot \phi}^*_k+\pi_k^*{\dot \phi}_k)
-\o_k^2(\phi^*_k\phi_k+\pi^*_k\pi_k)]\eqno(3.18)$$
Note that in the Fourier decomposition of the field $P(x)$ in (3.15), we had
intentionally incorporated  an additional factor of $k$ in front of $\pi_k$,
so that the form of (3.18) looks exactly like that of `complex'HO (2.18).
Also note that
the relation $\o_k^2=k^2$ was used crucially in these expressions, indicating
that the above structure for the Lagrangian is strictly 
valid for massless fields.

Mimicing the steps of the complex HO, it is simple to show that the
above Lagrangian displays either the $Z_2$ or $SO(2)$ symmetry, based on
a suitable relabelling of fields.

\vskip 0.5in

\noindent{\bf IV. Maxwell field in 4D}

\vskip 0.5in

In this section, we shall carry out a similar analysis for the free Maxwell
field. But because of the inherent gauge invariance of the model, we
shall not start with a Fourier analysis right at the beginning. Rather
the Gauss constraint of the model  will be
imposed strongly to isolate the physical degrees of freedom. Mode analysis
then reveals the HO structure just as in the scalar case with the difference
that to each mode ${\bf k}$ there are two orthogonal transverse oscillators.
Following the HO example,
this model is then linearised and written in `chiral' forms. We then carry out
the soldering of the `chiral' forms of the Lagrangian
followed by a summation over all the modes, to get hold of the final
soldered Lagrangian. This part is just the same as we did for the scalar
field. To that end, consider
$${\cal L}=-{1\over 4}F_{\mu \nu}F^{\mu \nu}={1\over 2}({\bf E}^2-{\bf B}^2)\eqno(4.1)$$
where $E_i=(\p_0A_i-\p_iA_0)$ and $B^k=\ep^{ijk}\p_iA_j$ are the electric
and magnetic fields. At this stage, the Gauss constraint 
(${\bf \nabla}.{\bf E}=0$) can be solved for $A_0$ yielding,
$$A_0={\p_0 \over {\nabla}^2}({\bf \nabla}.{\bf A})\eqno(4.2)$$
Time preservation of (4.2) is guaranteed by the equations of motion. Hence
it is possible to eliminate $A_0$ from the Lagrangian by using (4.2) to get
\cite{RB},
$${\cal L}={1\over 2}[({\dot {\bf A}^T})^2-
({\bf \nabla }\times {\bf A}^T)^2]\eqno(4.3)$$
where the longitudinal component ${\bf A}^L$ drops out automatically and
only the physical (transverse) component ${\bf A}^T$ survives. In terms of
the gauge field ${\bf A}$, this is given by,
$$A_i^T=P_{ij}A_j=(\d_{ij}-{\p_i\p_j\over {\bf {\nabla}}^2})A_j\eqno(4.4)$$
with $P_{ij}$ being the projection operator satisfying $P^2=P$. Therefore
the Fourier decomposition has to be carried out keeping this in mind.
Hereafter, we shall omit the superscript (T) from ${\bf A}^T$ and write
simply ${\bf A}$.

So finally carrying out a Fourier decomposition,
$${\bf A}(x)={1\over {\sqrt V}}\su e^{i{\bf k}.{\bf x}}
{\bf A}_k(t)\eqno(4.5)$$
Note that ${\bf A}_k(t)$ can be written as,
$${\bf A}_k(t)
=\sum_{\lambda=1}^2A_{k\lambda}(t){\bf{\ep}}_{\lambda}(k)\eqno(4.6)$$
where ${\bf{\ep}}_{\lambda}(k)$ are the polarisation vectors orthogonal
to ${\bf k}$
$({\bf k}.{\bf{\ep}}_{\lambda}(k)=0)$. This orthogonality projects out the
transverse component of the vector potential.

As expected, the Lagrangian $L(=\int d^3x {\cal L})$ can then be written
as,
$$L=\su L_k\eqno(4.7)$$
with 
$$L_k={1\over 2}({{\dot {\bf A}}_k^*}.{{\dot {\bf A}}_k}
-\o_k^2{{\bf A}_k^*}.{{\bf A}}_k)\eqno(4.8)$$
Comparing with (3.3b), we can see that ${\bf A}_k$ are now $3$-vectors
in ${\bf C}^3$ in contrast to $\phi_k$, which are just complex scalars.

Thus one can proceed just as in the scalar theory to linearise (4.8) by
invoking additional vector-valued variables ${\bf {\Pi}}_k$ and its complex
conjugates in an enlarged configuration space, to write
$$L_k={1\over 2}\o_k({\bf {\Pi}}_k^*.{\dot {\bf A}}_k +
{\bf {\Pi}}_k.{{\dot {\bf A}}_k^*})-{1\over 2}\o_k^2
({\bf {\Pi}}^*_k.{\bf {\Pi}}_k+{{\bf A}_k^*}.{\bf A}_k)\eqno(4.9)$$
associated with each mode $k$. Parametrising ${\bf q}_{1k}={\bf A}_k$
and ${\bf q}_{2k}={\bf {\Pi}}_k$ and then in the reverse order i.e.
${\bf q}_{2k}={\bf A}_k$ and ${\bf q}_{1k}={\bf {\Pi}}_k$, one gets
the following ``chiral'' forms of the Lagrangian,
$$L_{k\pm}({\bf Q}_k)={\pm}{1\over 2}\o_k{\bf Q}_k^{\dg}\ep .{\dot {\bf Q}}_k
-{1\over 2}\o_k^2{\bf Q}_k^{\dg}.{\bf Q}_k \eqno(4.10)$$
where ${\bf Q}_k$ is the doublet $\pmatrix{{\bf q}_{1k} \cr {\bf q}_{2k}}$.
The above lagrangian is invariant, mode by mode, under the usual $SO(2)$
transformation ${\bf Q}_k \rightarrow \ep {\bf Q}_k$. Similarly, under
the transformation  ${\bf Q}_k \rightarrow \s {\bf Q}_k$ the lagrangians
$L_{k+}$ and $L_{k-}$ are swapped into one another.

For a discussion of the equivalence of (4.10) with the standard form
\cite{R,SS,DGHT,BW} of duality invariant electromagnetic action, we refer
the reader to the appendix. The explicit  form of the duality generator
is also derived there. Alternatively, parametrising,
$${\bf {\Phi}}_{1k}= {\bf A}_k$$
$${\bf {\Phi}}_{2k}=-i{\bf {\Pi}}_k \eqno(4.11)$$
and then in the reverse order, the lagrangian (4.9) is expressed in the
chiral form as,
$$L_{k \pm}={1\over 2}(\pm i\o_k{\dot {\bf \Phi}}^{\dg}_k(t)\s {\bf \Phi}_k(t)
-\o_k^2{\bf \Phi}_k^{\dg}(t){\bf \Phi}_k(t))\eqno(4.12)$$
which reveals the $Z_2$ invariance, instead of the usual $SO(2)$. 
The analogy with the `complex' HO is therefore complete. Not surprisingly
therefore, a similar equation (3.18) had also occured earlier in the case
of scalar field. The only additional feature in this case is that
${\bf \Phi}_k= \pmatrix{{\bf \Phi}_{1k} \cr {\bf \Phi}_{2k}}$ is now a
doublet of vector fields.

It is quite straightforward to solder the two `chiral' forms of the
Lagrangians $L_{k+}({\bf Q}_k)$ and $L_{k-}({\bf R}_k)$ in the lines of the
scalar case to get,
$$L_k^s={1\over 2}({\dot {\bf S}}_k^{\dg}.{\dot {\bf S}}_k-\o_k^2
{\bf S}_k^{\dg}.{\bf S}_k)\eqno(4.13a)$$
where,
$${\bf S}_k={1\over {\sqrt 2}}({\bf Q}_k-{\bf R}_k)\eqno(4.13b)$$
is a doublet of {\it vectors}. Contrast this with (3.5b), where $S_k$
stands for a doublet of {\it scalars}.

Now to obtain the final soldered Lagrangian, we have to sum over all the
modes $(L^s=\su L_k^s)$, as we did for the scalar case (3.6). This yields,
$$L^s=-{1\over 4}\int d^3x G_{\mu \nu}^{\a}G^{\a \mu \nu}\eqno(4.14a)$$
where
$$G^{\a}_{\mu \nu}=\p_{\mu}A_{\nu}^{\a}-\p_{\nu}A_{\mu}^{\a}\eqno(4.14b)$$
is a doublet of abelian field strengths $(\a =1,2)$. This is the same result,
which was obtained in \cite{BW}. We have thus been able to provide an
slternative derivation by starting from the basic HO example.

Finally, introducing a doublet of divergence free vector field,
$${\bf S}(x)={1\over {\sqrt V}}\su e^{i{\bf k}.{\bf x}}{\bf S}_k(t)\eqno(4.15)$$
one can also cast $L^s$ in the pattern of scalar fields (3.7) as,
$$L^s={1\over 2}\int d^3x \p_{\mu}{\bf S}^\dg (x)\p^{\mu}{\bf S}(x)\eqno(4.16)$$
Again, the only difference  with (3.7) is that $S(x)$ appearing there
is a doublet of scalar fields, in contrast to the case here,
where ${\bf S}(x)$ is a doublet of vector fields.

The reason that the soldered Lagrangian for electrodynamics can be
cast in the form of scalars is rooted to the fact that, at the level of modes,
both of them represent an infinite number of decoupled
HOs. The only additional feature of electrodynamics is that to each mode 
${\bf k}$, there exists two orthogonal HOs associated to two polarisation
states.

\vskip 0.5in

\noindent{V.{\bf Conclusions}}

\vskip 0.5in

This paper showed that duality symmetry in certain free field theories had
their origin in a similar symmetry in a quantum mechanical example-the
`complex' harmonic oscillator(HO). While a clear distinction is made in the
literature concerning duality in $D=4k$ and $D=(4k+2)$ dimensions, nothing
specific is mentioned regarding the quantum mechanical case, which
can be regarded as a field theory in $D=1$ dimension. Our analysis, on the
other hand, clearly revealed that the study of duality symmetries in the
HO case is fundamental to properly understand the corresponding phenemenon
in the field theoretic case, at least for the models considered in this paper.
Indeed by performing an explicit mode analysis, the free scalar and Maxwell
theories were mapped to the complex HO. The one to one correspondence
between duality symmetry in the HO and the field theories was easily
established.

An algebraic consistency check was also provided for the mode analysis.
This was done by taking recourse to the soldering mechanism that was earlier
advocated by one of us \cite{BW,ABW}. It was shown that the soldering of duality
symmetric lagrangians (${\cal L}_+$ and ${\cal L}_-$) before the mode
decomposition yields identical results by, alternatively, first doing a
mode analysis of the individual lagrangians, then soldering the various modes
and finally summing over all the modes.

To understand the new feature in this paper it is necessary to recall the
development of duality symmetry. Originally, by considering the transformations
on the electric and magnetic fields, an invariance of the equations of motion
was found although the Lagrangean flipped its sign. In fact, as shown
here, this duality is obtained directly from the algebraic transformation
theory and need not consider any equations of motion. Obviously,
therefore, it was necessary to look at the invariance of the
action. Moreover, since the electric and magnetic fields are
derived quantities from the potential, it was reasonable to
study duality symmetry through these potentials. Simultaneously this brought
out a new feature, namely, the invariance of the action itself.
Nevertheless, a distinction between twice odd and twice even
dimensions prevailed since the duality groups in the two cases differed. 
By pushing this development to its logical
conclusion of considering the potential not as a basic field,
but as a quantity derived from its Fourier modes, and then investigating
duality symmetry through these modes, we obtained new results. The
invariance of the action was now demonstrated for both the
duality groups $Z_2$ and  $SO(2)$,
irrespective of the dimensionality of space time.
The explicit computations were done for the scalar theory
in $D=2$ dimensions and the Maxwell theory in $D=4$ dimensions.
Indeed, by  mapping these models
to the HO, it became clear that these have a common origin. By  suitable 
field redefinitions it was possible  to discuss the role of either $Z_2$
or $SO(2)$ as a duality group in both the models. The germ of this feature
was obviously contained in the HO, which displayed both the symmetries depending
on the change of variables. This may be compared with the general algebraic
arguments \cite{DGHT} regarding $Z_2$ and $SO(2)$ as the duality groups for $D=(4k+2)$
and $D=4k$ dimensions, respectively. Nevertheless, it should be clarified that there
is no clash between the general algebraic arguments and our findings. The duality
groups suggested by these arguments are obtained from the intersection of two groups
which, in turn, follow from the transformation properties of the N-form potentials. The crucial
point is that these forms are real whereas the modes with which we work are, in
general, complex. Hence such form based analysis does not cover our formalism. Yet another
way to look at this issue is to realise that the duality symmetry in field theories
were mapped to the corresponding properties in the H.O. The latter, being defined in
one dimension, obviously has no form related interpretation.
Clearly, therefore, our explicit
calculations provided fresh insights that cannot be otherwise gained from
purely general reasoning. We feel that this analysis of duality symmetry
through a mode expansion can be pursued for other examples.

\vskip 0.5in

\noindent{\bf Appendix}

\vskip 0.5cm

One may have observed that the entire discussion of duality symmetry for
the Maxwell theory did not invoke the familiar form of the Lagrangian
$${\cal L}_{\pm}={1\over 2}(\pm \ep_{\a \b}{\bf E}_{\a}.{\bf B}_{\b}
-{\bf B}_{\a}.{\bf B}_{\a})\eqno(A1a)$$
where,
$$E_{i \a}=\p_0A_{i \a}-\p_iA_{0 \a}$$
and
$$B_{i \a}=\ep^{ijk}\p_jA_{k \a} \eqno(A1b)$$
represent the electric and magnetic fields in the internal space.
This is further simplified to 
$${\cal L}_\pm={1\over 2}(\pm\ep_{\a \b}{\dot {\bf A}}_{\a}.{\bf B}_{\b}
-{\bf B}_{\a}.{\bf B}_{\a})\eqno(A2)$$
since the $A_0$ piece merely contributes a boundary term. The above form
of the duality invariant Lagrangian was obtained in various ways
\cite{DT,SS,BW} and has also been the starting point of several recent 
investigations \cite{B,KP,PST,G}. It is therefore desirable to establish
some sort of connection of our analysis with this structure.
Note that henceforth we  only
consider the positive `chiral' component of (A2) here.

Performing a mode analysis of (A2), we obtain, for the Lagrangian
$L(=\int d^3x {\cal L})$:
$$L=\su L_k \eqno(A3a)$$
where,
$$L_k={1\over 2}({\ep }_{\a \b}{\dot {\bf A}}_{\a k}^*.{\bf B}_{\b k}
-{\bf B}_{\a k}^*.{\bf B}_{\a k}) \eqno(A3b)$$
Using (A1b) and the fact that only the transverse components of the fields
are relevant, one finds the following relations,
$${\bf B}_{1k}=i{\bf k}\times {\bf A}_{1k}$$
$${\bf B}_{2k}=\o_k {\bf \Pi}_k \eqno(A4)$$
$${\bf A}_{2k}={i\over \o_k}{\bf k}\times {\bf \Pi}_k
={i\over \o_k^2}{\bf k}\times {\bf B}_{2k}$$
Inserting these in (A3) yields,
$$L_k={1\over 2}\o_k({\bf {\Pi}}_k^*.{\dot {\bf A}}_{1k} +
{\bf {\Pi}}_k.{{\dot {\bf A}}_{1k}^*})-{1\over 2}\o_k^2
({\bf {\Pi}}^*_k.{\bf {\Pi}}_k+{{\bf A}_{1k}^*}.{\bf A}_{1k})\eqno(A5)$$
which reproduces (4.9). This shows the equivalence of the duality invariant
Maxwell action derived here with the conventional form.

It is well known \cite{DGHT,BW} that the generator of the duality rotation
is given by the Chern-Simons structure,
$$
G=-\frac{1}{2}\int d^3 x {\bf A}_\a.{\bf B}_\a\eqno(A6)$$
Using the identification (A4), this reduces to,
$$
G_{k}=\frac{i}{2}{\bf{k}}.\Big({\bf{q}}_{\a k}^*\times {\bf{q}}_{\a k}\Big)
\eqno(A7)
$$
where ${\bf{q}}_{1 k}={\bf A}_k$ and ${\bf{q}}_{2 k}={\bf \Pi}_k$. This gives 
the explicit form of the generator corresponding to the $k -th$ mode (4.10).
It bears a striking resemblance with the expression obtained earlier for the
complex H.O.

We now show how
a real HO can also
be cast in the electromagnetic form. 
Noting that a one dimensional HO  can be regarded as a linear oscillator undergoing
its motion in a line passing through the origin and pointed in an
arbitrary direction  in ${\bf R}^3$, it is possible to write (2.1) as,
$$L={1\over 2}({\dot {\bf Q}}_1^2-\o^2{\bf Q}_1^2)=
{1\over 2}({\dot Q}_1^T{\dot Q}_1-\o^2Q_1^TQ_1)\eqno(A8)$$
where $Q_1=\pmatrix{q_1 \cr q_2 \cr q_3}\in {\bf R}^3$ is a triplet of real
numbers. Note that we have an enlarged 3-dimensional configuration space,
although the original HO undergoes its motion in 1-dimension.
In analogy with (A4), one can define,
$${\bf B}_1={\bf k}\times {\bf Q}_1 \eqno(A9)$$
where ${\bf k}$ is an arbitrary fixed vector in ${\bf R}^3$, which is
orthogonal to ${\bf Q}_1({\bf k}.{\bf Q}_1=0)$ and the magnitude is 
$\o=|{\bf k}|$. Similarly, define,
$${\bf B}_2={\dot {\bf Q}}_1\eqno(A10)$$
and
$${\bf Q}_2={1\over {\o^2}}{\bf k}\times {\bf B}_2\eqno(A11)$$
Note that ${\dot {\bf Q}_1}$ and ${\bf Q}_1$ are parallel. Since we have
assumed that ${\dot {\bf k}}=0$, we can easily see that the vectors
$({\bf Q}_1,{\bf Q}_2,{\bf k})$ form an orthogonal triplet.
Then (A8) can be written as,
$$L={1\over 2}(\ep_{\a \b}{\dot {\bf Q}}_{\a}.{\bf B}_{\b}
-{\bf B}_{\a}.{\bf B}_{\a})\eqno(A12)$$
which maps with the usual duality invariant form (A2) of the Maxwell theory.


\newpage

\begin{thebibliography}{99}
\bibitem{R} For recent reviews, see D.I. Olive, Exact Electromagnetic 
Duality, hep-th /9508089; Nucl. Phys. {B58} (Proc. Suppl.), 43 (1997);
L. Alvarez-Gaume and F. Zamora, Duality in Quantum Field Theory
(string theory), hep-th/9709180.
\bibitem{SS} J.Schwarz and A.Sen, Nucl.Phys.{\bf B411},35(1994).
\bibitem{DGHT} S.Deser, 
A.Gomberoff, M.Henneaux and C.Teitelboim, Phys.Lett.{\bf B400}
80(1997).
\bibitem{BW} R.Banerjee and 
C.Wotzasek,``Bosonisation and duality symmetry in the soldering formalism'',
hep-th/9805109,To appear in Nucl.Phys.{\bf B}.
\bibitem{S} M.Stone, Illinois preprint  ILL-(TH)-89-23 (1989); Phys.Rev.Lett.
{\bf 63}, 731 (1989); Nucl.Phys.{\bf B327}, 399(1989).
\bibitem{ADW} R.Amorim, A.Das and C.Wotzasek, Phys.Rev.{\bf D53},5810(1996).
\bibitem{ABW} 
E.M.C. Abreu, R.Banerjee and C.Wotzasek, Nucl.Phys.{\bf B509}519,(1998).
\bibitem{RB} R.Banerjee, Report No. hep-th/9610240.
\bibitem{DT} S.Deser and C.Teitelboim, Phys.Rev.{\bf D13}, 1592(1976).
\bibitem{B} N.Berkovits, Phys.Lett.{\bf B388}, 743(1996).
\bibitem{KP} A.Khoudeir and N.Pantoja, Phys.Rev. {\bf D53}, 5974(1996).
\bibitem{PST} P.Pasti, D.Sorokin and M.Tonin, Phys.Lett. {\bf B352},59(1995), 
Phys.Rev.{\bf D52},R4277(1995).
\bibitem{G} H.O.Girotti, Phys.Rev. {\bf D55},5136(1997).

\end{thebibliography}
\end{document}